\documentclass[traditabstract, rnote]{aa}
\usepackage[utf8]{inputenc}
\usepackage{setspace}
\usepackage{amsmath}
\usepackage{amssymb}
\usepackage{microtype}
\usepackage{natbib}
\usepackage{graphicx}
\usepackage{txfonts}
\usepackage{color}
\usepackage{multirow}
\usepackage{xspace}
\bibpunct{(}{)}{;}{a}{}{,} 
\newcommand{\ero}{\emph{eRosita}\xspace}
\newcommand{\ig}{\emph{INTEGRAL}\xspace}
\newcommand{\es}{\,erg\,s$^{-1}$\xspace}
\newcommand{\escm}{\,erg\,s$^{-1}$\,cm$^{-2}$\xspace}

\title{Population of the Galactic X-ray binaries and \ero}
\author{V.\,Doroshenko, L.\,Ducci, A.\,Santangelo, M.\,Sasaki}
\institute{Institut für Astronomie und Astrophysik, Sand 1, 72076 Tübingen, Germany}

\begin{document}

\bibliographystyle{aa}

\abstract{The population of the Galactic X-ray binaries has been
mostly probed with moderately sensitive hard X-ray surveys so far.
The \ero mission will provide, for the first time a sensitive all-sky
X-ray survey in the 2-10\,keV energy range, where the X-ray
binaries emit most of the flux and discover the still unobserved
low-luminosity population of these objects. In this paper, we briefly
review the current constraints for the X-ray luminosity functions of
high- and low-mass X-ray binaries and present our own analysis based
the \ig 9-year Galactic survey, which yields improved constraints.
Based on these results, we estimate the number of new XRBs to
be detected in the \ero all-sky survey.}

\authorrunning{V. Doroshenko et al.}
\maketitle

\section{Introduction} The evolution of astrophysical objects usually occurs on
timescales that are inaccessible to direct observations ($\ge10^6$\,yr), and
population studies are the only way to tackle it. From an observational point
of view, the distribution of sources with luminosity or the luminosity function
(LF) is often used to characterise a given population. Since the X-ray binaries
(XRB) emit most of the energy in X-rays, the X-ray luminosity function (XLF) is
usually considered. It has been shown that the XLF of both low and high mass
X-ray binaries (LMXB, HMXB) is a broken power law flattening below
$\sim10^{36}-10^{37}$\es \citep[and references therein]{Grimm:2002hv,
Voss:2010ft, Mineo2011, Lutovinov:2013dt}. Taking the power-law
nature of the XLF into account, most of the sources are expected to have low luminosities,
so observing them is essential for constraining the basic properties of XRB
population. With current instrumentation, this can only be done for the Galactic
XRBs based on the data from all-sky X-ray surveys \citep{Grimm:2002hv}.

Highly non-uniform spatial distribution of XRBs within the Milky Way, however,
makes it non-trivial to reconstruct the XLF from the observed fluxes and assess
the completeness of a given survey. The non-uniform X-ray absorption further
complicates the situation, especially for soft X-ray surveys like \emph{ROSAT},
so currently hard X-ray surveys by \emph{Swift} \citep{Ajello:2008em} and
\emph{INTEGRAL} \citep{Krivonos:2012il} observatories provide the best
constraints on the XLF parameters. However, both surveys are limited in
sensitivity to $\sim10^{-11}$\escm. Therefore, probing the XLF below $10^{35}$\es
is still challenging.

The \emph{Swift BAT} survey is being carried out in the 15-55\,keV energy
range, and the majority of the sources detected so far are X-ray binaries. For
sources with known distances, \cite{Voss:2010ft} derived the observed XLF and
corrected it for being incomplete by following the approach suggested by
\cite{Grimm:2002hv}. For the LMXBs, the authors confirm the previously reported
\citep{Grimm:2002hv,Revnivtsev:2008iw} break in the XLF at
$\sim3\times10^{36}$\es with slopes of $\sim0.9$ and $\sim2.4$ below and above
the break, respectively. For the first time, they found that the XLF of HMXBs
also likely has a break at $L_{br}\sim2.5\times10^{37}$\es with slopes of $\sim
1.3$ and $\ge2$ below and above the break, respectively.

\cite{Lutovinov:2013dt} investigated the XLF of HMXBs detected in the
\emph{INTEGRAL} 9-year survey in the 17-60\,keV energy range considering only
the persistent sources. To assess the completeness of the sample,
\cite{Lutovinov:2013dt} divided it into a set of flux-limited samples within
concentric annuli around the Galactic centre based on known distances. The
surface density of HXMBs was assumed constant within each annulus, while the XLF
shape was assumed to be the same throughout the Galaxy. The corrected XLF was
found to have a break at $2.5\times10^{36}$\es and slopes of $\sim1.4$
and $\ge2.2$ for the dim and bright parts respectively, which is consistent
with the values reported by \cite{Voss:2010ft}.

The reported uncertanties for XLF parameters are, however, rather large in both
cases, which makes it difficult to extrapolate results to lower luminosities
and to estimate the total number of yet undetected XRBs in the Milky Way. The
uncertanties are mainly driven by the limited size of the source sample, particularly
at low luminosities. Indeed, limited sensitivity of \ig and \emph{Swift BAT}
surveys implies that weaker sources may only be detected if they are
close to Earth, and these are not many.

The upcoming \emph{eRosita} all-sky X-ray survey \citep{Merloni:2012ug} will
bring a significant improvement in sensitivity over the existing surveys and is
likely to detect a number of new low-luminosity XRBs thus improving the
constraints on the XLF parameters. The main objective of the present paper is to
assess the anticipated impact of \emph{eRosita} survey in this context.

To get a more robust forecast, we start by analysing the XLF
of the Galactic XRBs using available data, namely the \ig 9-year Galactic plane
survey \citep{Krivonos:2012il}. It is the most sensitive survey available for
low Galactic latitudes, where most XRBs are located, and the resulting source
catalogue contains about twice as many identified XRBs as the \emph{Swift BAT}
survey used by \cite{Voss:2010ft}. To further increase the size of the
source sample, we propose a novel method for XLF reconstruction, which does not
require the knowledge of the distances to individual sources but rather relies
on modeling the observed flux distribution of a given survey based on the input
model XLF and the spatial distribution of the sources in the sample. The
results are consistent with the values reported by
\cite{Voss:2010ft} and \cite{Lutovinov:2013dt}, although we do find a flatter
low-luminosity slope and a lower break luminosity (and, therefore, fewer
objects) for the HMXBs. Using these results, we then conclude optimistically that
the \emph{eRosita} survey might be expected to double the number
of the observed XRBs in Galaxy and improve the constraints on the
XLF parameters if sufficient identification completeness is
achieved.

\section{XLF reconstruction} The standard method to characterise the luminosity
distribution of a population is to calculate it directly from the observed
fluxes for a sub-sample of objects with known distances. The observed XLF must then
be corrected for effects of incompleteness that are related, for instance, to
the inhomogeneous survey sensitivity or spatial distribution of the population
members. For instance, the LMXBs roughly trace the stellar mass distribution, and
the sources located in the Galactic bulge represent substantial fraction of
entire population. However, only relatively luminous bulge LMXBs can be
detected in shallow surveys, so a substantial fraction of all low-luminosity
LMXBs in Galaxy are not detected and, therefore, the observed XLF are
biased towards high-luminosity sources. The observed XLF can be corrected for
this effect by estimating (for each luminosity) the fraction of sources
expected to be detected in a survey, providing that the source spatial
distribution is known or can be modeled \citep{Grimm:2002hv}.

The corrected luminosity distribution depends, however, on not only
flux and distance estimates for the observed sources but also on the
overall distribution of distances to sources in population. We
suggest, however, that the expected flux distribution for any assumed
intrinsic luminosity distribution in population can be calculated
with no additional assumptions in this case. The parameters of the
assumed XLF might then be tuned so that the model and observed flux
distributions match, and therefore, the intrinsic XLF of the
population may be recovered. The reconstructed XLF obviously
depends on the assumed spatial distribution of sources in population,
but this is also the case for the directly derived XLF due to
incompleteness effects. Modeling of the observed flux rather than
luminosity distribution has, however, an important advantage that
even the sources lacking distance estimates might be included in the
analysis thus reducing statistical uncertanties. We note that
the observed flux or luminosity distribution is still affected by
identification completeness effects which must be considered
separately in XLF reconstruction.

A very similar approach has been used by \cite{pretorius07}
to assess the influence of selection effects on the observed
population of cataclysmic variables (CVs) in the optical domain.
These authors also modeled the flux distribution of the observed CV
population by assuming certain intrinsic luminosity distribution,
spatial distribution of the CVs, their spectral properties, and
outburst activity. We note that other effects such as dependence of
the identification completeness in a survey on flux might also be
accounted. The latter is likely to be important for the upcoming \ero
survey. However, it is not trivial to assess how complete
identification is before the actual survey is finished. Now, we are
more interested in calculating the total number of XRBs likely to be
detected by \ero. To assess the expected improvement in constraints
on XLF parameters, we optimistically assume that most sources will be
eventually identified.

To make a more realistic forecast for \ero and improve the
constraints on XLF parameters using available data, we first
reconstruct the XLF using the \ig 9-year Galactic
survey catalogue \citep{Krivonos:2012il}. This provides survey
averaged fluxes in the 17-60\,keV energy range for 108 LMXBs and 82
HMXBs. This dataset has very high identification completeness
and is not sensitive to interstellar absorption, which makes it ideal
for population studies. Therefore, it can be safely assumed that the
observed flux distribution is only defined by the intrinsic XLF,
spatial distribution of the sources, and the sensitivity of the
survey. Moreover, this dataset has not been used for detailed
analysis of the XLF of X-ray binaries so far.

Only a fraction of the total flux is emitted in 17-60\,keV, so the
simulated bolometric flux needs to be multiplied by some factor to
derive the flux detected by a particular instrument. To estimate the
X-ray contribution detected by \emph{INTEGRAL}, we calculate
$F_{17-60}/F_{1-200}$ ratios explicitly using the spectral parameters
reported for some of the known XRBs in the \ig reference catalogue
\citep{Ebisawa:2003fe} and randomly pick the conversion factor from
the resulting empirical distribution for the simulated sources. This
is particularly important for LMXBs, which exhibit well-defined
hard/soft spectral states and are hard to describe by just stacking
spectra of all sources to derive the hard X-ray contribution, as done
by \cite{Voss:2010ft}.

The sensitivity of a survey in a given direction is assumed to depend
only on exposure as $F_{lim}\sim1.1\times10^{-8}/\sqrt{T_{exp}}$\escm
\citep{Krivonos:2012il} and can be calculated from the survey
exposure maps for each simulated source to determine whether it would
be detected in the survey.

\begin{figure*}[tb]
	\centering
		\includegraphics[width=0.49\textwidth]{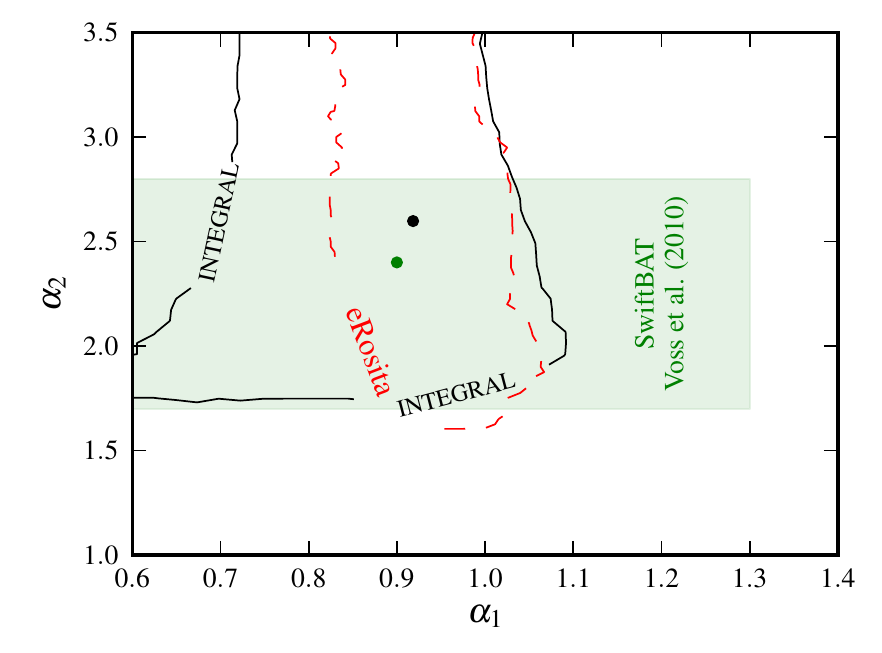}
		\includegraphics[width=0.49\textwidth]{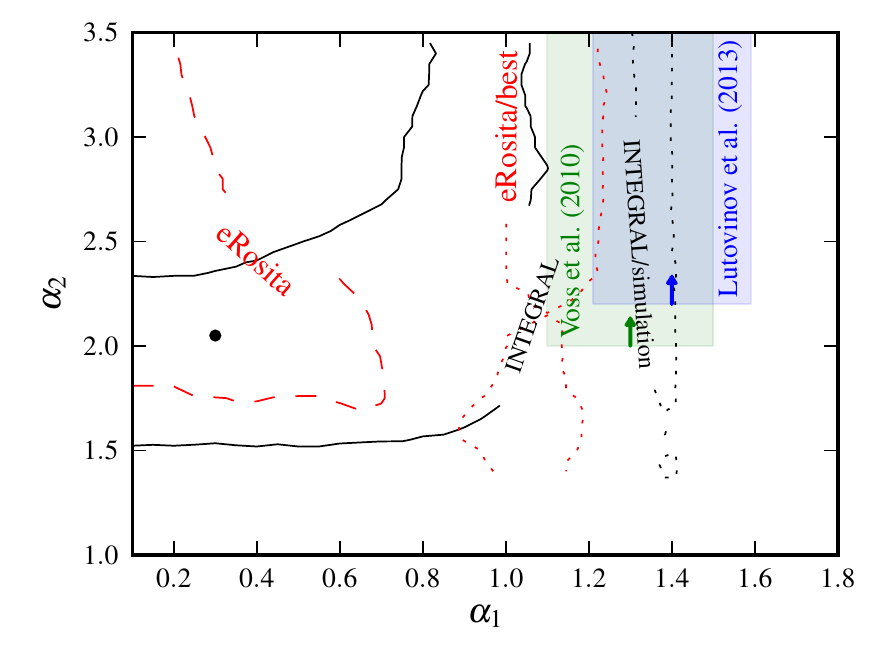}
	\caption{Simulation results for the LMXB (left) and the HMXB (right)
populations. Black contours and points show a 68\% probability that the
simulated population will have the same flux distribution as observed in \ig
9-year Galactic survey and the best-fit results. The red countours show the
forecast for the \ero 4-year survey assuming the best-fit XLF found from \ig
data.
The black and red dotted contours represent results for the synthetic HMXB
populations 1) by assuming the XLF parameters reported by \cite{Lutovinov:2013dt} and
observed with \ig and 2) by assuming the
best-fit XLF parameters with $\alpha_1=1.1$ fixed at the highest value allowed and
observed with \ero, which is an optimistic forecast.
The XLF parameters as reported by \cite{Voss:2010ft} and \cite{Lutovinov:2013dt}
are also shown for reference.}
	\label{fig:XLF}
\end{figure*}
\begin{figure*}[tb]
	\centering
		\includegraphics[width=0.49\textwidth]{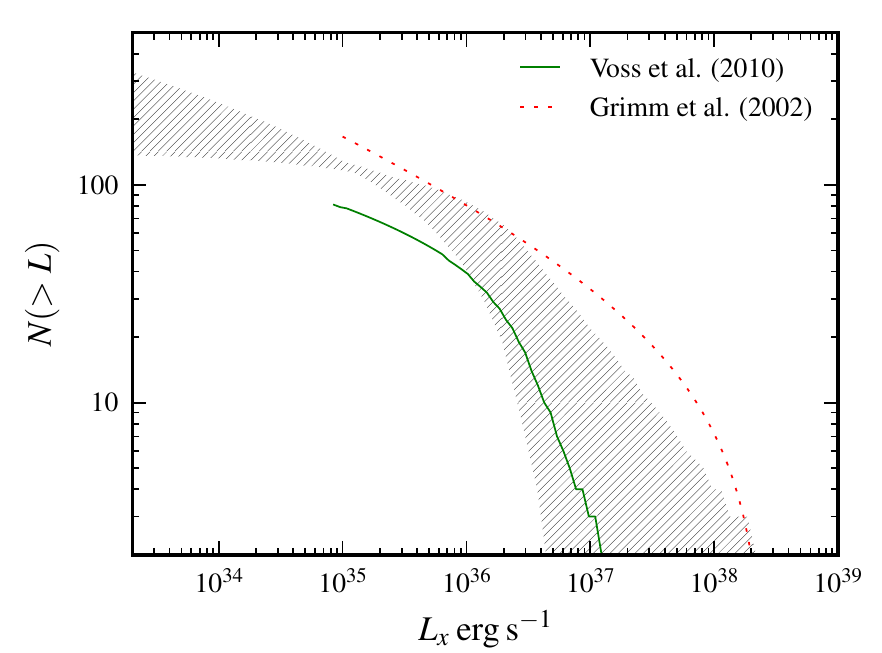}
		\includegraphics[width=0.49\textwidth]{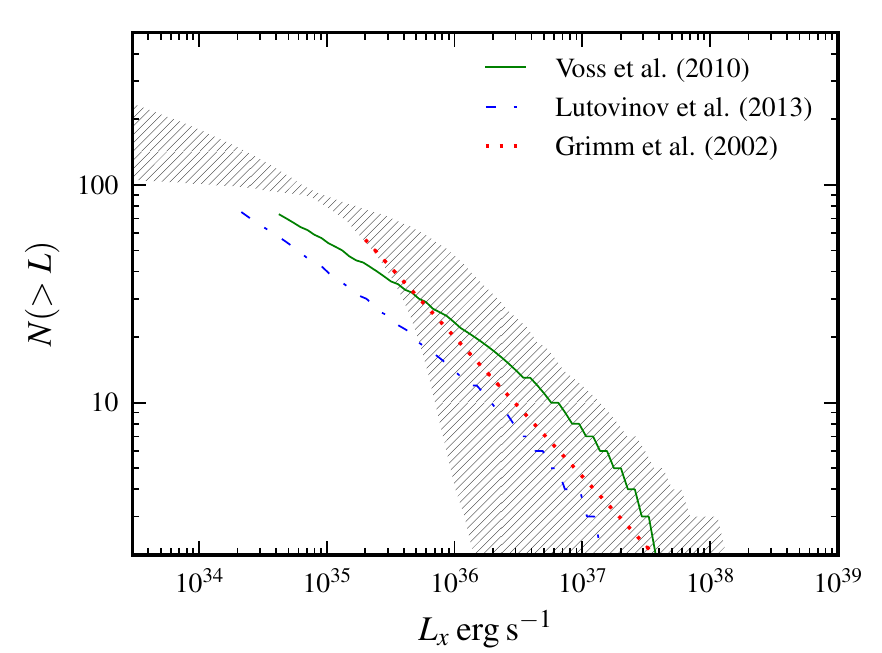}
	\caption{Cumulative luminosity functions for LMXBs (left) and HMXBs (right)
	as derived in this work using the \ig data (hatched area). Best-fit estimates by \cite{Grimm:2002hv},
	\cite{Voss:2010ft}, and \cite{Lutovinov:2013dt} are also plotted for reference.}
	\label{fig:XLF_cum}
\end{figure*}

To model the populations of HMXBs and LMXBs we use the spatial distributions of
XRBs obtained by \cite{Bahcall:1980ch} and \cite{Grimm:2002hv}, and for the spiral
arm shape, we use the parameters reported by \cite{Cordes:2002tt}. The XLF is
parametrised with a broken power-law
\hbox{$dN/dL\propto(L/L_{br})^{-\alpha_{1,2}}$} with three free parameters: the
break luminosity $L_{br}$ and the slopes $\alpha_{1}$ and $\alpha_2$ below and
above the break luminosity respectively. To obtain the observable flux
distribution, we simulate a population of objects with a given spatial and
intrinsic luminosity distributions, converting the luminosity to observed flux.
This is done for a set of input XLF parameters and then compared with the
observed flux distribution using the non-parametric two-sample
Kolmogorov-Smirnov test \citep{conover:1971}. This test gives the probability
$P_{\rm KS}$ that both the observed and simulated flux samples originate from
the same distribution, that is, the model XLF is able to reproduce the
observed fluxes. Setting a threshold on $P_{\rm KS}$ allows, therefore, the
restriction of the XLF parameters.

Note that the normalisation of the XLF only affects the
total number of observed sources but does not affect the shape of the
distribution and $P_{\rm KS}$ value. Therefore, it is not considered a free
parameter, and we calculate it later by comparing the number of sources predicted
for a given XLF with the total number of observed sources.
Our method to derive the XLF can be, therefore, summarised as follows:
\begin{itemize}
	\item Based on the spatial distribution model, we generate coordinates for
a large ($\sim10^{4}$) population of sources.
	\item We assume a broken power law XLF parametrized with $L_{br}, \alpha_1, \alpha_2$, and pick a random luminosity for each source from this model distribution;
	\item using the sampled coordinates and luminosity, we then calculate the expected bolometric flux
from each source, and convert it to instrumental flux using the spectral
information available from the \ig reference catalogue.
	\item For each simulated source we determine whether it would be detected by \ig by comparing its flux with the survey sensitivity in
a given direction and obtain the final model flux distribution.
	\item We calculate the normalisation of the model XLF by comparing
	the number of sources detected in simulation with number of sources detected in the \ig survey.
	\item Finally, we compare the model and the observed flux distributions at each point
on a grid of input XLF parameters $L_{br}, \alpha_1, \alpha_2$ using the
two-sample Kolmogorov-Smirnov test to calculate the probability $P_{\rm KS}(L_{br},
\alpha_1, \alpha_2)$ that two samples come from the same distribution. Setting
a threshold on $P_{\rm KS}$ allows the acceptable values for XLF
parameters to be restricted.
\end{itemize}
To smooth out the variations due to the random nature of simulation, we repeat
the steps above about ten times for each combination of XLF parameters until the
resulting sample of $P_{\rm KS}$ values become normally distributed and then calculate
the mean value.

\section{Results and discussion} Taking the previous estimates of
the XLF parameters into account, we considered the parameter
ranges of $0.1\le\alpha_1\le1.8$, $0.9\le\alpha_1\le\alpha_2\le3.5$, and
$10^{35}\le L_{br}\le 10^{38}$. The results are presented in
Table.~\ref{tab:main} and Fig.~\ref{fig:XLF}. Here, the contours mark regions
where the maximum value of $P_{\rm KS}$ for a given $\alpha_{1,2}$ and any
$L_{br}$ exceeds the selected threshold of $P_{\rm KS}\ge68$\%. Earlier
reported estimates of XLF parameters by \cite{Voss:2010ft, Lutovinov:2013dt}
are also shown for reference. The respective cumulative luminosity functions
are shown in Fig.~\ref{fig:XLF_cum}.

\begin{table}[t]
	\begin{center}
		\begin{onehalfspace}
	\begin{tabular}{lcc}
		\hline
		& LMXB & HMXB \\
		\hline
	$L_{br},10^{36}$\,\es & 8$_{-6.5}^{+7}$ & 0.55$_{-0.28}^{+4.6}$ \\
	$\alpha_1$ & $0.9_{-0.4}^{+0.2}$ & $0.3_{-0.2}^{+0.8}$ 	\\
	$\alpha_2$ & $2.6_{-0.9}^{+3}$ & $2.1_{-0.6}^{+3}$ 	\\
	$N_{total, MW}$ & $200_{-75}^{+175}$ & $110_{-10}^{+180}$ \\
	$N_{INTEGRAL}$ & 108 (29-86\%) & 82 (27-82\%)\\
	$N_{eRosita}$ &  130-270 (75-95\%) & 105-220 (78-96\%)\\
	$N_{eRosita,new}$ & 22-162 & 23-138 \\
	\hline
	\end{tabular}
	\end{onehalfspace}
	\end{center}
	\caption{Estimated XLF parameters for high- and low- mass X-ray binaries from \ig 9-year survey,
	and expected number and fraction of sources to be detected by \ig and \ero.}
	\label{tab:main}
\end{table}

Our results for both the HMXBs and LMXBs are consistent within uncertanties
with previous reports. We find, however, a somewhat flatter slope at low
luminosities and a lower break luminosity than both \cite{Voss:2010ft} and
\cite{Lutovinov:2013dt}. However, being based on a different (extended) source
sample and a different XLF reconstruction method, our analysis is completely
independent, hence, some differences are expected. To verify the robustness of
our method, we repeated the analysis using a synthetic HMXB distribution,
which is derived by assuming the XLF parameters reported by \cite{Lutovinov:2013dt} as
an observational reference. We were able to reliably reconstruct the input
parameters (see Fig.~\ref{fig:XLF}). We conclude, therefore, that the slope of
HMXBs XLF found in previously mentioned works was likely overestimated. On the
other hand, our estimate might be also biased. Indeed, the simple estimate of
\ig sensitivity, which we use might be too optimistic at lower fluxes,
particularly if weak sources are observed within the same field of view with
brighter ones \citep{Fenimore78}. Some of the dim sources, which we assume
are detectable by \ig are, therefore, likely not detected. On the other hand,
the \ero survey will be free from such complications.

To get a forecast for the \ero survey, we use the same approach described
above. We simulate the expected XRB flux distributions using the expected
survey sensitivity and flux conversion factors calculated in a same way as for
\ig. Although the \ero is most sensitive to soft X-rays below $\sim2$\,keV,
the X-ray spectra of XRBs are known to be intrinsically hard and
often strongly absorbed, so we restrict our analysis to the hard 2-10\,keV
energy range where the absorption is less important.

For the input XLF parameters and normalisations, we take our results
from the analysis of the \ig data. The limiting flux for \ero in
2-10\,keV energy range is assumed to be
$F_{lim}\sim6.9\times10^{-12}/\sqrt{T_{exp}}$\escm where the exposure
maps are taken from \cite{Merloni:2012ug}. The results are presented
in Fig.~\ref{fig:XLF} and Table~\ref{tab:main}.

The \ig survey is already sensitive enough to detect most if not all sources
with high luminosities, so the uncertainty in the XLF slope above the break for
both HMXBs and LMXBs is mainly driven by the limited number of such sources in
Galaxy.

At low luminosities, \ero will detect a number of new sources, potentially
doubling the numbers of known XRBs. This does not, however, directly imply that
the constraints on the XLF parameters is significantly improved (see
Fig.~\ref{fig:XLF}). For the LMXBs, the uncertanties for $L_{br}$ and
$\alpha_1$ might be reduced by factor of two if all detected LMXBs are
identified. However, this is unlikely to be the case, as the source
identification is a major challenge for low-luminosity LMXBs, which are
easily confused with cataclysmic variables \citep{Jonker11, Torres:2013tl}.

For HMXBs, the source identification is less complicated, as their
luminous optical counterparts are detectable throughout the Galaxy
\citep{Reed03}. Therefore, \ero is likely to put a stronger limit on
$\alpha_1$ (see Fig.~\ref{fig:XLF}).

\section{Conclusions} We presented a detailed analysis of the X-ray luminosity
function of the Galactic X-ray binaries. Our analysis is based on the most
sensitive hard X-ray survey available, the \ig 9-yr Galactic Survey. Previously,
this dataset had only been used to constrain the XLF of persistent HMXBs,
whereas we consider the entire populations of both LMXBs and HMXBs. 

We also suggested a method for XLF reconstruction, which allowed us
to further increase the source sample size by the inclusion of
sources with no distance estimates. We model the observed flux rather
than luminosity distribution and considering the spectral properties
of individual objects and the variable survey sensitivity accross the
sky. Our results generally agree with the previous reports, although
we find a flatter slope for HMXBs at low luminosities, which implies
that there are fewer HMXBs in the Milky Way. Based on the
reconstructed XLF parameters and using a similar simulation
procedure, we conclude that \ero shall detect up to 160 new LMXBs and
140 HMXBs and significantly improve the constraints on the low
luminosity part of the XLF of the Galactic X-ray binaries. We note,
however, that such analysis requires survey identification to be
reasonably complete, which will be particularly challenging for the
LMXBs.

\begin{acknowledgements}
authors thank the Deutsches Zentrums für Luft- und Raumfahrt (DLR) and Deutsche
Forschungsgemeinschaft (DFG) for financial support (grants DLR~50~OR~0702, FKZ
50 OG 1301, SA2131/1-1).
\end{acknowledgements}

\bibliography{auto_clean}

\begin{thebibliography}{17}
\expandafter\ifx\csname natexlab\endcsname\relax\def\natexlab#1{#1}\fi

\bibitem[{Ajello {et~al.}(2008)Ajello, Greiner, Kanbach, Rau, Strong, \&
  Kennea}]{Ajello:2008em}
Ajello, M., Greiner, J., Kanbach, G., {et~al.} 2008, ApJ, 678, 102

\bibitem[{Bahcall \& Soneira(1980)}]{Bahcall:1980ch}
Bahcall, J.~N. \& Soneira, R.~M. 1980, ApJS, 44, 73

\bibitem[{Conover(1971)}]{conover:1971}
Conover, W.~J. 1971, Practical nonparametric statistics

\bibitem[{Cordes \& Lazio(2002)}]{Cordes:2002tt}
Cordes, J.~M. \& Lazio, T. J.~W. 2002

\bibitem[{Ebisawa {et~al.}(2003)Ebisawa, Bourban, Bodaghee, Mowlavi, \&
  Courvoisier}]{Ebisawa:2003fe}
Ebisawa, K., Bourban, G., Bodaghee, A., Mowlavi, N., \& Courvoisier, T. J.-L.
  2003, A{\&}A, 411, L59

\bibitem[{{Fenimore} \& {Cannon}(1978)}]{Fenimore78}
{Fenimore}, E.~E. \& {Cannon}, T.~M. 1978, \ao, 17, 337

\bibitem[{Grimm {et~al.}(2002)Grimm, Gilfanov, \& Sunyaev}]{Grimm:2002hv}
Grimm, H.~J., Gilfanov, M., \& Sunyaev, R. 2002, MNRAS, 391, 923

\bibitem[{{Jonker} {et~al.}(2011){Jonker}, {Bassa}, {Nelemans}, {Steeghs},
  {Torres}, {Maccarone}, {Hynes}, {Greiss}, {Clem}, {Dieball}, {Mikles},
  {Britt}, {Gossen}, {Collazzi}, {Wijnands}, {In't Zand}, {M{\'e}ndez}, {Rea},
  {Kuulkers}, {Ratti}, {van Haaften}, {Heinke}, {{\"O}zel}, {Groot}, \&
  {Verbunt}}]{Jonker11}
{Jonker}, P.~G., {Bassa}, C.~G., {Nelemans}, G., {et~al.} 2011, \apjs, 194, 18

\bibitem[{Krivonos {et~al.}(2012)Krivonos, Tsygankov, Lutovinov, Revnivtsev,
  Churazov, \& Sunyaev}]{Krivonos:2012il}
Krivonos, R., Tsygankov, S., Lutovinov, A., {et~al.} 2012, A{\&}A, 545, A27

\bibitem[{Lutovinov {et~al.}(2013)Lutovinov, Revnivtsev, Tsygankov, \&
  Krivonos}]{Lutovinov:2013dt}
Lutovinov, A.~A., Revnivtsev, M.~G., Tsygankov, S.~S., \& Krivonos, R.~A. 2013,
  MNRAS, 431, 327

\bibitem[{Merloni {et~al.}(2012)Merloni, Predehl, Becker, B{\"o}hringer,
  Boller, Brunner, Brusa, Dennerl, Freyberg, Friedrich, Georgakakis, Haberl,
  Hasinger, Meidinger, Mohr, Nandra, Rau, Reiprich, Robrade, Salvato,
  Santangelo, Sasaki, Schwope, Wilms, \& German~eROSITA
  Consortium}]{Merloni:2012ug}
Merloni, A., Predehl, P., Becker, W., {et~al.} 2012, eprint arXiv:1209.3114

\bibitem[{{Mineo} {et~al.}(2011){Mineo}, {Gilfanov}, \& {Sunyaev}}]{Mineo2011}
{Mineo}, S., {Gilfanov}, M., \& {Sunyaev}, R. 2011, Astronomische Nachrichten,
  332, 349

\bibitem[{{Pretorius} {et~al.}(2007){Pretorius}, {Knigge}, \&
  {Kolb}}]{pretorius07}
{Pretorius}, M.~L., {Knigge}, C., \& {Kolb}, U. 2007, \mnras, 374, 1495

\bibitem[{{Reed}(2003)}]{Reed03}
{Reed}, B.~C. 2003, \aj, 125, 2531

\bibitem[{Revnivtsev {et~al.}(2008)Revnivtsev, Lutovinov, Churazov, Sazonov,
  Gilfanov, Grebenev, \& Sunyaev}]{Revnivtsev:2008iw}
Revnivtsev, M., Lutovinov, A., Churazov, E., {et~al.} 2008, MNRAS, 491, 209

\bibitem[{Torres {et~al.}(2013)Torres, Jonker, Britt, Johnson, Hynes, Greiss,
  Steeghs, Maccarone, Ozel, Bassa, \& Nelemans}]{Torres:2013tl}
Torres, M. A.~P., Jonker, P.~G., Britt, C.~T., {et~al.} 2013

\bibitem[{Voss \& Ajello(2010)}]{Voss:2010ft}
Voss, R. \& Ajello, M. 2010, ApJ, 721, 1843

\end{thebibliography}
\end{document}